\documentclass[twocolumn,showpacs,preprintnumbers,amsmath,amssymb,floatfix]{revtex4}

\usepackage{graphicx}
\usepackage{dcolumn}
\usepackage{bm}
\usepackage{color}
\usepackage{epstopdf}

\newcount\timehh  \newcount\timemm
\timehh=\time \divide\timehh by 60 \timemm=\time
\definecolor{gold}{rgb}{0.85,.66,0}

\begin{document}

\title{All-optical NMR in semiconductors provided by resonant cooling
of nuclear spins interacting with electrons in the resonant spin amplification regime }

\author{E.~A.~Zhukov,$^{1}$ A.~Greilich,$^{1}$ D.~R.~Yakovlev,$^{1,2}$
K.~V.~Kavokin,$^{2,3}$ I.~A. Yugova,$^{1,3}$
O.~A.~Yugov,$^{1,3}$ D.~Suter,$^{4}$ G.~Karczewski,$^{5}$ T.~Wojtowicz,$^{5}$
J.~Kossut,$^{5}$ V. V. Petrov,$^{3}$ Yu. K. Dolgikh,$^{3}$ A.~Pawlis,$^{6}$  and M. Bayer$^{1}$}

\affiliation{$^1$ Experimentelle Physik 2, Technische Universit\"at Dortmund, 44221 Dortmund, Germany}
\affiliation{$^2$ Ioffe Physical-Technical Institute, Russian Academy of Sciences, 194021 St. Petersburg, Russia}
\affiliation{$^3$ Physical Faculty of Saint Petersburg State University, 198504 St. Petersburg, Russia}
\affiliation{$^4$ Experimentelle Physik 3, Technische Universit\"at Dortmund, 44221 Dortmund, Germany}
\affiliation{$^5$ Institute of Physics, Polish Academy of Sciences, 02668 Warsaw, Poland}
\affiliation{$^6$ Department Physik, Universit\"at Paderborn, 33098 Paderborn, Germany}

\date{\today, file = \jobname.tex, printing time = \number\timehh\,:\,\ifnum\timemm<10 0\fi \number\timemm}

\begin{abstract}
Resonant cooling of different nuclear isotopes manifested in
optically-induced nuclear magnetic resonances (NMR) is observed in
\textit{n}-doped CdTe/(Cd,Mg)Te and ZnSe/(Zn,Mg)Se quantum wells and
for donor-bound electrons in ZnSe:F and GaAs epilayers. By
time-resolved Kerr rotation used in the regime of resonant spin
amplification we can expand the range of magnetic fields where the
effect can be observed up to nuclear Larmor frequencies of 170~kHz.
The mechanism of the resonant cooling of the nuclear spin system is
analyzed theoretically. The developed approach allows us to model
the resonant spin amplification signals with NMR resonances.
\end{abstract}

\maketitle

\section{Introduction}\label{sec:intro}

Nuclear magnetic resonance (NMR) is a well-established technique
which is widely used to analyze structures and electronic states in
solids \cite{Abragam1961,Ernst1997}. NMR is one of the key
technologies for the implementation of quantum information processing, as the
nuclear spins are almost ideal qubits for the
manipulation and storage of the quantum information
\cite{Nielsen2000}. The realization of this potential  in semiconductor nanostructures
requires significant technical improvements to reach extremely high sensitivity and
nanometer-scale resolution.

An important step in this direction was the
optical detection of NMR (ODNMR) excited by radio-frequency  fields.
For this purpose, the effect of resonant changes of the nuclear polarization on  the electron spin
polarization can be measured through the polarization of the photoluminescence
\cite{Ekimov1972,Berkovits1973,Dyakonov74,Paget1981,Kalevich1983,Eickhoff2002}
or as Faraday and Kerr rotation \cite{Kennedy2006}, for reviews see
Refs.~\cite{Kalevich_Ch11,OptOr-5,OptOr-9}. Being well established for bulk
semiconductors, ODNMR has been
successfully applied also to semiconductor quantum wells (QWs)
\cite{Kalevich90, Flinn90, Eickhoff03, Eickhoff04, Poggio05, Sanada06,Kondo08} and quantum dots (QDs)
\cite{Gammon1997} with much smaller numbers of  nuclear
spins contributing to the signal.

The next key achievement was the realization of the optically-induced
NMR or the so-called ``all-optical NMR'', where the dynamical nuclear
polarization (DNP) was induced and detected by purely optical means.
For the resonant addressing of the NMR the rf magnetic field is
replaced with the oscillating Knight field of the spin-polarized
electrons \cite{Kalevich1986,Kalevich88,Eickhoff02}. The oscillating
Knight field is provided by either intensity or polarization
modulation of the laser light that photogenerates spin-oriented
electrons in semiconductors. All-optical NMR has also been
realized on the basis of the time-resolved pump-probe Faraday/Kerr
rotation technique, where the coherent Larmor precession of the
electron spins is
detected~\cite{Kik00,Sal01a,Sal01b,Awschalom_Spintronics}. NMR has
been demonstrated in different magnetic fields and corresponding Larmor frequencies
by using different techniques for the modulation of the laser light:
mechanical choppers ($1-6$~kHz), photo-elastic modulators
($50-100$~kHz), electro-optical modulators ($1-10$~MHz),
or the repetition frequency of mode-locked lasers of typically around
80~MHz.

The all-optical NMR technique can manifest itself in two different
ways: resonant heating or resonant cooling of the nuclear spin
system (NSS). The resonant heating is typically observed for pumping
light with fixed circular polarization, i.e. under conditions when
the spin polarization flows from the  oriented carriers to the NSS,
which provides efficient DNP and reduces the nuclear spin
temperature. When the pumping light is modulated in intensity and
the modulation frequency coincides with the NMR frequency of some
nuclear isotope, resonant heating of the NSS occurs.  In a typical
experiment, the Zeeman splittings of the nuclei are tuned by
external magnetic field and sharp NMR features are observed when a
resonance condition is met.

Resonant cooling is observed when  pumping light with
alternating helicity is used. Under these conditions, the dynamic
nuclear polarization  is strongly suppressed if the modulation
period is shorter than the transverse relaxation time in the
NSS~\cite{OptOr}. However, if the modulation frequency matches the
NMR frequency in the external magnetic field applied in the Voigt geometry,
the efficiency of the DNP sharply increases resulting in the
resonant cooling of the NSS. The reason for the DNP enhancement is
synchronization of the Larmor precession of the injected
non-equilibrium nuclear spin with the oscillating Knight field,
which results in an efficient extraction of entropy from the NSS and therefore
a reduction of the spin temperature.
The nuclear polarization gained in this way has so far been
observed via the contribution of the Overhauser field to the
Hanle effect, which is the  depolarization of the electron spin in
the transverse magnetic field \cite{Kalevich80, Kalevich95}.
The signal from the resonant cooling has a typical dispersion-like shape, as distinct from the resonant heating that produces absorption-like signals.

For semiconductor nanostructures, the all-optical NMR by means
of the pump-probe Faraday/Kerr rotation has been realized on
modulation-doped GaAs/(Al,Ga)As quantum wells under conditions of
the resonant heating of the NSS ~\cite{Kik00,Sal01a,Sal01b}. We are not aware about observation of the resonant cooling of the NSS by these techniques.

The use of the Hanle effect for the NMR detection imposes
limitations on the range of applicable magnetic fields and,
consequently, of NMR frequencies. These limitations are especially
severe for structures with long electron spin lifetimes, where Hanle
curves are as narrow as a few Gauss. The corresponding NMR
frequencies do not exceed 10~kHz, which is comparable with the width
of the resonance lines in semiconductors and, therefore, does not
allow to resolve important details like isotopic structure,
quadrupole splitting, etc. This situation is typical for
\textit{n}-doped quantum wells and quantum dots. Electron spin
dephasing times in these structures can be as long as $30-100$~ns,
which has been documented for quantum wells based on GaAs, CdTe and
ZnSe semiconductors
\cite{Astakhov2008,Zhukov2014,Griesbeck,rsa_paper}. It exceeds the
typical repetition period of the mode-locked pulsed lasers of 13~ns
and an accumulation of the electron spin coherence can be realized
here. The resulting electron spin polarization can be conveniently
measured in the resonant spin amplification (RSA) regime
~\cite{Kikkawa98,Yugova12}.

In this paper we extend the all-optical NMR studied for the resonant
spin amplification regime of the pump-probe Kerr rotation technique
and investigate a variety of II-VI QWs and II-VI and III-V
semiconductor epilayers. The chosen experimental conditions result
in sharp NMR features in the RSA spectra associated with the
resonant cooling of the nuclear spin system. We develop a
theoretical approach to analyze the underlying mechanism, which
predicts further modifications of the RSA spectra due to dynamical
nuclear polarization.

The paper is organized as follows. Section~\ref{sec:1} provides details of the
experimental techniques and of the studied samples.
Section \ref{sec:exp} describes the experimental results.
Section~\ref{sec:theory} is devoted to  quantitative theoretical
considerations of the resonant nuclear-spin cooling in
\textit{n}-doped quantum wells. Experimental results are compared with the modeling.
In the conclusions, we compare our RSA technique for studying the nuclear spin system
with earlier experimental techniques.

\section{Experimentals}
\label{sec:1}

Time-resolved pump-probe Kerr rotation (TRKR) technique in the
resonant spin amplification (RSA) regime
~\cite{Awschalom_Spintronics, Kikkawa98,Yugova12} was used to study
the interaction of electron spins with the nuclear spin system and
demonstrate the resonant nuclear-spin cooling. The electron spin
coherence was generated by a train of  circularly polarized pump
pulses of 1.5~ps duration (spectral width of about 1~meV) generated
by  a mode-locked Ti:Sapphire laser operating at a repetition
frequency of 75.7~MHz (repetition period $T_R=13.2$~ns). The pump
helicity was modulated between $\sigma^+$ and $\sigma^-$
polarizations by means of photo-elastic modulators (PEM) operating
at frequencies of $f_\mathrm{m} = 42$, 50, and 84~kHz, so that in
average the samples were equally exposed to both polarizations of
the pump. To alternate the circular light with 84~kHz we used the
the PEM in $\lambda/4$ retardation mode and set the lock-in
amplifier to this frequency. In order to double the operation
frequency up to 168~kHz, this PEM was used in a $\lambda/2$
retardation mode with an additional $\lambda/4$ plate placed at an
angle  of $45^{0}$ to the linear polarization axis and the lock-in
detection was set to the second harmonic.

The photon energy of the pump pulse was tuned into resonance with
the negatively charged excitons (negative trions, T$^-$) of the
studied QWs, which allows us to generate spin coherence for the
resident electrons in the QWs~\cite{zhu07,chapter6}. To realize that
in ZnSe-based QWs the pump beam was frequency doubled by a nonlinear
BBO crystal ~\cite{Zhukov2014}. For the epilayers the laser photon
energy was resonant with the donor-bound exciton optical
transitions. The induced electron spin coherence (either for the
resident electron in QWs or for the electrons on donors) was
monitored by linearly polarized probe pulses reflected from the
excited area, which were time delayed with respect to the pump
pulses by a mechanical delay line. The rotation angle of the
polarization plane of the reflected probe beam (Kerr rotation) was
measured by a balanced photodetector connected to a lock-in
amplifier. The pump and probe beams had the same photon energy. For
the RSA measurements the probe pulse arrival time was fixed at a
small negative delay ($\Delta t \sim -50$~ps) prior to the pump
pulse and the magnetic field was scanned across a small range close
to zero. The average pump power was kept at the relatively low level
of $P_{\mathrm{pump}}=1-5$~W/cm$^2$, and the probe power
($P_{\mathrm{probe}}$) was about one order of magnitude smaller than
that of the pump.

The samples were placed in a vector magnet system consisting of
three orthogonal superconducting split-coils \cite{Zhukov12}. The
measurements were performed in magnetic fields up to 3~T applied
perpendicular to the structure growth axis,
$\mathbf{B}\perp\mathbf{z}$, and for the pump wave vector
($\mathbf{k}_{\mathrm{pump}}$) parallel to the \textit{z}-axis, i.e.
in Voigt geometry. The vector magnet system allows us to compensate
residual magnetic fields along other  axes, which are commonly
present in superconducting split-coil solenoids. Samples were
immersed in pumped liquid helium at a temperature of $T=1.8$~K.

For photoluminescence (PL) measurements a continuous-wave (cw) laser
with photon energy of 2.32~eV was used for the CdTe-based QWs and
GaAs epilayer, and a cw laser with 3.05~eV for the ZnSe-based
structures. PL signals were detected with a Si-based
charged-coupled-device camera attached to an 0.5-m spectrometer.

For this study we selected samples with sufficiently long spin
dephasing times $T_2^*>30$~ns in order to be able to measure them in
the RSA regime. Quantum well structures and epilayers based on three
different material systems of the II-VI and III-V semiconductors
CdTe/(Cd,Mg)Te, ZnSe/(Zn,Mg)Se, ZnSe:F and GaAs  were used.

The studied CdTe/Cd$_{0.78}$Mg$_{0.22}$Te QW heterostructure
(\#031901C, sample\#1) was grown by molecular-beam epitaxy on an
(100)-oriented GaAs substrate followed by a 2 $\mu$m CdTe buffer
layer. It has five periods, each of them consisting of a 20-nm-thick
CdTe QW and a 110-nm-thick Cd$_{0.78}$Mg$_{0.22}$Te barrier. An
additional 110-nm-thick barrier was grown on top of this layer
sequence to reduce the contribution of surface charges. The barriers
were modulation doped with iodine donors. Electrons from the barrier
donors, being collected in the QWs, provide there a two-dimensional
electron gas (2DEG) with a density of about
$n_{e}=1.1\times10^{10}$~cm$^{-2}$. Detailed studies of  the optical
properties and the carrier spin coherence in this structure were
published in  Refs.~\cite{zhu06,zhu07,Astakhov2008}. The $g$-factor
of the resident electrons, $g_{e} = -1.64\pm 0.02$ was determined
from the Larmor precession frequency in magnetic fields exceeding
0.5~T.

A homogeneously fluorine-doped 100-nm-thick ZnSe epilayer was grown
by molecular-beam epitaxy on $(001)$-oriented GaAs substrate
(\#2029, sample \#2).  The concentration of the fluorine donors  is
about $10^{15}$~cm$^{-3}$. The ZnSe:F layer was grown on top of a
20-nm-thick Zn$_{1-x}$Mg$_{x}$Se buffer layer that prevents carrier
diffusion into the GaAs substrate. The magnesium concentration of
this layer was kept below 15\% to maintain good crystal quality. The
optical properties of this sample and information on the electron
spin coherence can be found in Ref.~\cite{Greilich12}. The
$g$-factor  for the donor-bound electrons is $1.13\pm 0.02$.

A 20-nm-thick ZnSe/Zn$_{1-x}$Mg$_{x}$Se single QW  was grown by
molecular-beam epitaxy on a $(001)$-oriented GaAs substrate (\#2018,
sample \#3). The QW is surrounded by Zn$_{1-x}$Mg$_{x}$Se barrier
layers with thicknesses of 24 and 30~nm. The magnesium concentration
of these layers was kept below 15\%. This structure was nominally
undoped, but due to residual impurities and charge redistribution to
surface states the QW at low temperatures contains resident
electrons with density not exceeding $10^{10}$~cm$^{-2}$.  The
$g$-factor for the resident electrons in the QW  is $1.13\pm 0.02$.

The studied GaAs epilayer was grown by molecular-beam epitaxy on a
$(001)$-oriented semi-insulating GaAs substrate (sample \#4). The
epilayer was nominally undoped,  but contains residual donors with concentration of about $10^{16}$~cm$^{-3}$ and $g_{e} = -0.44\pm 0.02$.

\section{Experimental results}
\label{sec:exp}

\subsection{CdTe-based QW}
\label{sec:2}

\begin{figure}[hbt]
\includegraphics*[width=8.5cm]{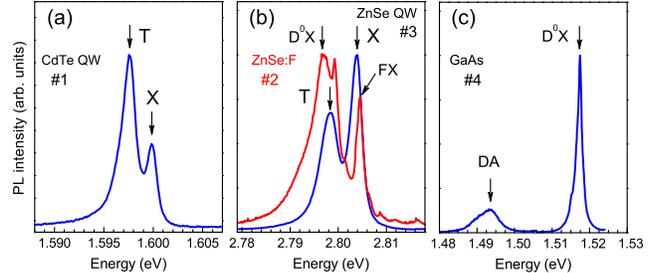}
\caption[] {(Color online) Photoluminescence spectra of studied
structures: (a) 20-nm-thick CdTe/(Cd,Mg)Te QW; (b) ZnSe:F epilayer
and 20-nm-thick ZnSe/(Zn,Mg)Se QW; (c) GaAs epilayer. Exciton (X),
trion (T), donor-bound exciton D$^{0}$X, free exciton (FX) and
donor-acceptor recombination (DA) lines are marked by arrows.
$B=0$~T and $T=1.8$~K. } \label{Fig:1}
\end{figure}

\begin{figure}[hbt]
\includegraphics*[width=8.5cm]{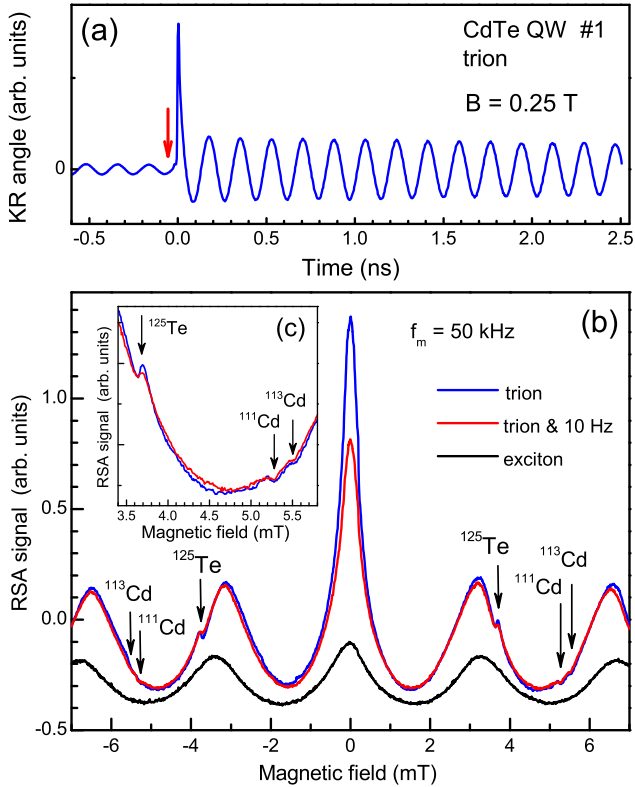}
\caption[] {(Color online) (a) Kerr rotation pump-probe signal of
20-nm-thick CdTe/(Cd,Mg)Te QW measured at trion resonance.
$B=0.25$~T and $T=1.8$~K. Red arrow shows the time delay of $\Delta
t = -80$~ps at which the RSA signals were detected. (b) RSA signals
measured at the trion (blue line) and exciton (black line)
resonances for $f_\mathrm{m} = 50$~kHz.  The red line shows the
trion signal measured with an  additional amplitude modulation of
the pump beam at 10~Hz. $P_{\mathrm{pump}}=1$~W/cm$^2$ and
$P_{\mathrm{probe}}=0.2$~W/cm$^2$. Insert (c) zooms  the  NMR
features. Calculated $B_{\mathrm{NMR}}$ from Table I for different
isotopes are marked by arrows. } \label{Fig:2}
\end{figure}

Figure~\ref{Fig:1}(a) shows the photoluminescence spectrum of the
20-nm-thick CdTe/(Cd,Mg)Te QW, which consists of the exciton and
trion emission lines separated by the trion binding energy of 2~meV.
The pump-probe Kerr rotation signal measured for this QW on the
trion resonance at $B=0.25$~T is shown in Fig.~\ref{Fig:2}(a).  The
characteristic oscillations correspond to the Larmor precession of
the electron spin about the external magnetic field. This signal is
observed at much longer delays than the exciton or trion
recombination times, which are shorter than 100~ps \cite{zhu07}.
Therefore, it can be related to the spin coherence of resident
electrons. The dephasing of this spin coherence lasts longer than
the pump repetition period $T_R=13.2$~ns. As a result, a
considerable signal amplitude is observed at negative delays, i.e.
prior the pump pulse arrival. This  behavior is qualitatively
similar for all samples studied in this paper. Therefore, for other
samples we will show only RSA spectra.

The long dephasing time $T_2^*$ of the electron spin coherence
exceeding $T_R=13.2$~ns allows us to perform measurements in the RSA
regime.  Figure~\ref{Fig:2}(b) shows the RSA signals for the trion
and exciton resonances with the magnetic field scanned from $-7$ to
$+7$~mT. A typical RSA signal has periodically spread RSA peaks. The
peak distance on the magnetic field scale corresponds to one period
of the electron Larmor precession. The width of the RSA peaks is
determined by the spin dephasing time $T_2^*$
~\cite{Kikkawa98,Yugova12}.

The unusual feature of the RSA signals from the trion, shown by the
blue curve in Fig.~\ref{Fig:2}(b), is the strong intensity of the
zero-field RSA peak with  respect to the other peaks at  finite
magnetic fields. An  additional amplitude modulation of the pump
beam at a very low frequency of 10~Hz by a mechanical chopper
resulted in suppression of  the  zero field peak amplitude by 40\%,
as shown by the red curve. This is a strong hint that the
enhancement of the zero-field RSA peak is related to the
polarization of the nuclear spin system and a feedback of the nuclei
on the electron spin polarization measured by RSA. The enhancement
is almost  absent for the RSA signals measured on the excitons (not
shown). This is in line with the suggested explanation, as in the
case of trion the localized resident electrons can better polarize
nuclei in the  volume of their localization, compared to the weaker
localized electrons, whose  spin coherence is induced via excitons.
It was shown for CdTe/(Cd,Mg)Te QWs that resident electrons with
different localization are addressed via the trion and exciton
states, see Fig.~20 in Ref.~\onlinecite{zhu07}.

Interaction between the electron and nuclear spin systems appears
also in the form of relatively weak dispersive resonance features in
the RSA signals at magnetic fields of $\pm 3.7; \pm 5.3$ and $\pm
5.6$~mT, where the NMR resonances at $f_\mathrm{m} = 50$~kHz for
$^{125}$Te, $^{111}$Cd and $^{113}$Cd isotopes are expected. The
arrows in Fig.~\ref{Fig:2}(b) show the calculated resonance magnetic
fields $B_{\mathrm{NMR}}$ for these isotopes, the numerical values
are collected in Table~I. Similar NMR resonances were reported
earlier for Hanle curves of III-V semiconductor structures  being
related to the resonant cooling of the NSS
\cite{Kalevich88,Kalevich95}. Since these experiments were performed
under  different experimental conditions, one needs to develop a
suitable theory that can describe resonant cooling of the NSS and
its detection in the RSA regime. Such a model will be suggested
below.

Note that in order to observe NMR features in RSA signals a finite
signal amplitude is required. I.e., the NMR resonances are very weak
and hardly observable between RSA peaks, where the signal amplitude
has a  minimum. To make them more pronounced one can either vary the
modulation frequency $f_\mathrm{m}$ to shift $B_{\mathrm{NMR}}$ or
broaden the RSA peaks by shortening the dephasing time, e.g. by
means of increasing the pump power.

\begin{table} [b]
\caption{Calculated values of $B_{\mathrm{NMR}}$ given in mT for the
resonant features experimentally observed in the studied structures
~\cite{Landoldt,Whit10,syperek11,merkulov02,Nakamura79,Paget77}.
$I_{\alpha}$ is the nuclear spin and $\mu_{\alpha}$ is the nuclear
magnetic moment for the specific isotope $\alpha$.} \label{Table 1}
\begin{tabular}
{p{1.1cm}p{0.9cm}p{1.2cm}p{1.1cm}p{1.1cm}p{1.1cm}p{1.1cm}}
\hline
\\& & &\multicolumn{4}{c}{Resonance magnetic field (mT)} \\
\hline
\\Isotope&$I_{\alpha}$&$\mu_{\alpha}$ &42~kHz&50~kHz&84~kHz&168~kHz \\
\hline
\\$^{111}$Cd &1/2 &$-$0.5943 & &5.52& & \\
\\$^{113}$Cd &1/2 &$-$0.6217 & &5.28& & \\
\\$^{125}$Te &1/2 &$-$0.8871 & &3.69& & \\
\\$ ^{67}$Zn &5/2 &+0.8754 &  &18.67& & \\
\\$ ^{77}$Se &1/2 &+0.534 &  &6.14&10.3&20.7\\
\\$ ^{69}$Ga &3/2 &+2.016 &4.10&  & &  \\
\\$ ^{71}$Ga &3/2 &+2.562 &3.23&  & &  \\
\\$ ^{75}$As &3/2 &+1.439 &5.75&  & &  \\
\hline
\end{tabular}
\end{table}

\subsection{ZnSe:F epilayer and ZnSe-based QW }\label{sec:3}

Photoluminescence spectra of the ZnSe:F epilayer and ZnSe/(Zn,Mg)Se
QW are shown in Fig.~\ref{Fig:1}(b). The epilayer spectrum consists
of several lines (for details see Ref.~\onlinecite{Greilich12}), two
of them relevant to this study are marked by arrows. They are the
donor-bound exciton (D$^{0}$X) at 2.7970~eV and free exciton with
heavy hole (FX) at 2.8045~eV. The binding energy of the exciton to
the  fluorine donor is about 7.5~meV. The QW spectrum  has two lines
at 2.7984 and 2.8030~eV corresponding to the trion and exciton
recombination, respectively, in the ZnSe QW. They are separated by
4.6~meV, which is the binding energy of the negatively charged
trion.

\begin{figure}[hbt]
\includegraphics*[width=8.5cm]{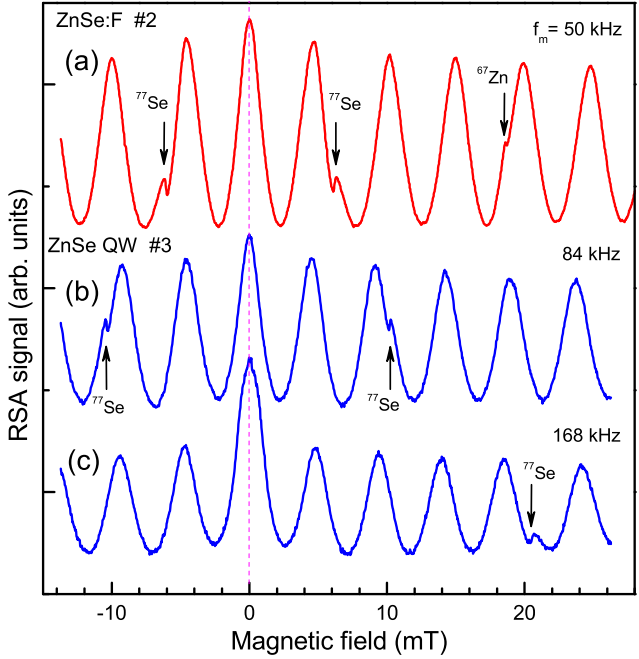}
\caption[] {(Color online) RSA signals for ZnSe:F epilayer (a) and
20-nm-thick ZnSe/(Zn,Mg)Se QW (b,c) measured  for different  modulation
frequencies at $T=1.8$~K. $\Delta t = -35$~ps,
$P_{\mathrm{pump}}=5$~W/cm$^2$ and $P_{\mathrm{probe}}=0.2$~W/cm$^2$.
Calculated $B_{\mathrm{NMR}}$ from Table I are shown by arrows.  }
\label{Fig:3}
\end{figure}

Examples for the pump-probe Kerr rotation signals in the ZnSe:F
epilayers and ZnSe-based QWs can be found in
Refs.~\onlinecite{Greilich12,Zhukov2014}. The RSA signals for these
samples, measured for various modulation frequencies are shown in
Fig.~\ref{Fig:3}. As one can see, the RSA spectra of the epilayer
and QW are very similar to each other. Pronounced NMR resonances for
the $^{77}$Se and $^{67}$Zn isotopes are clearly seen. They are
shifted to higher magnetic fields with increasing modulation
frequency from 50 up to 168~kHz.  For $f_\mathrm{m}= 50$~kHz
resonances are seen at $\pm 6.2$ ($ ^{77}$Se) and $+18.6$~mT ($
^{67}$Zn), for $f_\mathrm{m}=84$~kHz at $\pm 10.2$~mT ($ ^{77}$Se),
and for $f_\mathrm{m}=168$~kHz  at $ +20.5$~mT ($ ^{77}$Se).

\subsection{GaAs-epilayer}\label{sec:4}

\begin{figure}[hbt]
\includegraphics*[width=8cm]{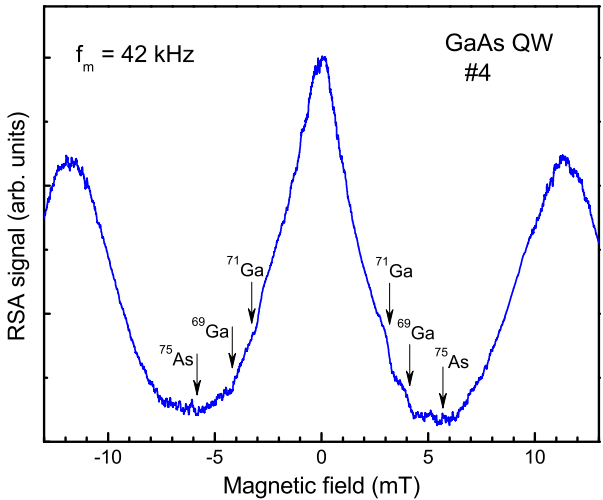}
\caption[] {RSA signal for GaAs epilayer measured at resonant
pumping of the  D$^{0}$X transition at $f_\mathrm{m}=42$~kHz.
$T=1.8$~K. $\Delta t=-50$~ps, $P_{\mathrm{pump}}=5$~W/cm$^2$ and
$P_{\mathrm{probe}}=0.2$~W/cm$^2$. Calculated $B_{\mathrm{NMR}}$
from Table I are shown by arrows. } \label{Fig:4}
\end{figure}

The photoluminescence spectrum of the GaAs epilayer (sample \#4) is
shown in Fig.~\ref{Fig:1}(c). It contains two lines at  1.5174~eV
and 1.4932~eV corresponding to the donor-bound exciton (D$^{0}$X)
and donor-acceptor recombination (DA), respectively~\cite{Landoldt}.
The RSA signal measured with resonant pumping of  the  D$^{0}$X
transition is shown in Fig.~\ref{Fig:4}. The in-plane electron
$g$-factor of $g_e=-0.44$ has been evaluated. NMR features for the
$^{71}$Ga, $ ^{69}$Ga and $ ^{75}$As isotopes are seen at negative
and positive magnetic fields of $\pm 3.2$, $\pm 4.1$ and $\pm
5.8$~mT. These fields match well the expected NMR resonances for
$f_\mathrm{m}=42$~kHz, compared with the calculated values  in Table
I.

\section{Theory}\label{sec:theory}

It has been shown in the experiment that the NMR features became
very pronounced in the RSA spectra under conditions, in which on the
first view no mean spin from the electrons should be transferred
into the NSS: (i) high symmetry of the structures with zinc-blende
crystal lattice grown along the (001)-axis; (ii) the pump beam is
parallel to the structure growth axis, $\mathbf{k}_{\mathrm{pump}} \parallel
\mathbf{z}$; (iii) the external magnetic field is perpendicular to
the structure growth axis,  $\mathbf{B} \perp \mathbf{z}$ and
$\mathbf{B} \perp \mathbf{k}_{\mathrm{pump}}$; (iv) the  structure  is
equally exposed by $\sigma^+$ and $\sigma^-$ circularly polarized
pump. To disclose the responsible mechanisms a detailed theoretical
analysis is needed, which is given in this Section.

\subsection{Cooling of the nuclear spin system via interaction
with spin polarized electrons}
\label{sec:5}

As mentioned in the introduction, the resonant cooling of the
nuclear spin system (NSS) in exact Voigt geometry is the  result of
synchronization of the spin injection into the  NSS, the oscillatory
magnetic field applied to the NSS (this can be the Knight field of
the electrons), and the Larmor precession of nuclear spins in the
static transverse magnetic field. In the case of RSA experiments,
this picture becomes more complicated, since pulsed excitation
brings in high-frequency components of the mean electron  spin,
which are essential for the overall spin dynamics, as their
superposition forms the observed series of RSA peaks. Here we extend
and develop the theory of resonant cooling ~\cite{OptOr-5} to adopt
it to this new experimental design.

The resonant cooling of the NSS can be theoretically described using
modified Provotorov equations for two effective temperatures
describing the dipole-dipole and Zeeman nuclear spin reservoirs
{~\cite{OptOr-5, Merkulov82}. In order to calculate the average
nuclear field and its effect upon the mean electron spin (which we
optically probe in RSA experiments), a more intuitive, though quite
rigorous, approach can be used, which invokes the rotating frame
approximation ~\cite{Kalevich88, OptOr-5}.

Under the conditions of RSA, DNP in the rotating frame has certain
specifics because of the presence of ``slow'' and ``fast''
oscillating components of the electron spin. Oscillation of the slow
component with an angular frequency $\omega_\mathrm{m} = 2 \pi
f_\mathrm{m}= (2.64 \div 10.55) \times 10^5$~rad/s is provided by
the PEM with $f_\mathrm{m} = 42 \div 168$~kHz, while the fast
component is due to electron spin Larmor precession. Under RSA
conditions the main contributions to the fast component are made by
oscillations with frequencies commensurate with the repetition rate
of the pulsed laser of 75.7~MHz. At the magnetic field corresponding
to the maximum of the first RSA peak the corresponding Larmor
frequency is $\omega_L= 2 \pi/T_\mathrm{R}=4.75 \times 10^8$~rad/s =
$2 \pi \times 75.7$~MHz. We will show below, that the slow component
of the electron spin polarization is responsible for the DNP, while
the resultant nuclear field is detected by its effect on the fast
components of the electron spin polarization.

We now consider the electron spin polarization photogenerated by the
pump pulses and averaged over many pulses and Larmor periods, as it
evolves as a function of the phase of the PEM. The average spin
polarisation is perpendicular to the external magnetic field
$\mathbf{B}$ and oscillates with the angular frequency
$\omega_\mathrm{m}$ of the PEM. The linear oscillation can be
represented as a superposition of two rotating vectors, which are
also perpendicular to $\mathbf{B}$:
\begin{equation}
\label{nuclear_gen}
\mathbf{S}(t) = \mathbf{S}_+(t)+ \mathbf{S}_-(t).
\end{equation}
Each of these mean-spin vectors produces a double effect on the NSS:
(i) spin relaxation of electrons by nuclei creates a spin flow into
the NSS, and (ii) the nuclear spins become subjected to a rotating
Knight field. If we turn to the rotating frame, where $\mathbf{S}_+$
(or $\mathbf{S}_- )$  is static, then $\mathbf{S}_-$ (or,
correspondingly, $\mathbf{S}_+)$ rotates at double frequency
$2\omega_\mathrm{m} \gg T^{-1}_{2,N}$ and its contribution to DNP
can be neglected. Here $T_{2,N}$ is the transverse spin relaxation
time of the nuclei. Under these conditions, the averaged (over the
rotation period) spin flow $\mathbf{q}$ is parallel to $\mathbf{B}$
and equals the sum of the contributions from $\mathbf{S}_+$ and
$\mathbf{S}_-$ ($\mathbf{q}_+$ and $\mathbf{q}_-$),  which can be
calculated independently in the frame where the corresponding spin
component is static:
\begin{eqnarray}
\label{nuclear_gen1}
q = q_+ + q_- &=& \frac{1}{T^{(e)}_{1,N}}
\frac{S_0}{2} \sum_{\alpha} \left[\frac{B_{e,\alpha} \tilde B_{+,\alpha}}{(\tilde B_{+,\alpha})^2 + (B_{e,\alpha})^2 + \xi
B_L^2} \right. \nonumber \\
&+& \left. \frac{B_{e,\alpha} \tilde B_{-,\alpha}}{(\tilde
B_{-,\alpha})^2 + (B_{e,\alpha})^2 + \xi B_L^2} \right].
\end{eqnarray}
Here $S_0$ is the averaged electron spin, $T^{(e)}_{1,N}$ is the
longitudinal spin relaxation time of the nuclei due to their
interaction with electrons,  $\tilde B_{\pm,\alpha} =B  \pm
\gamma^{-1}_{\alpha} \omega_\mathrm{m}$ (the term $\pm
\gamma^{-1}_{\alpha} \omega_\mathrm{m}$ along $B$ occurs at the
transition to the rotating frame), $\gamma_{\alpha}$ is the
gyromagnetic ratio  of the $\alpha$th isotope,
$B_{e,\alpha}=b_{e,\alpha} S_0 /2 $ is the Knight field,
$b_{e,\alpha}=-A_{\alpha}/(\hbar\gamma_{\alpha})$, $A_{\alpha}$ is
the hyperfine constant for the isotope $\alpha$, $B_L$ is the rms
local field due to the nuclear dipole-dipole interaction, and $\xi$
is a dimensionless parameter of the order of one, accounting for
electron spin correlation ~\cite{OptOr-5}. We refer the reader to
Ref.~\cite{OptOr-5} for a detailed theory of nuclear spin cooling in
the Knight field leading to Eq.~\eqref{nuclear_gen1}.

\begin{figure}[hbt]
\includegraphics*[width=8cm]{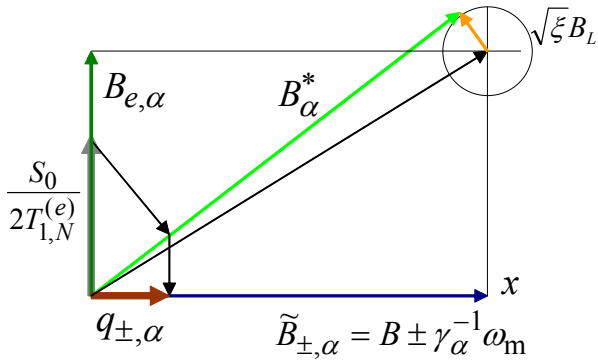}
\caption[] {Scheme of the electron-nuclear spin interaction
process in the rotating coordinate system.}
\label{Fig:5}
\end{figure}

The electron-nuclear spin interaction processes are shown
schematically in Fig.~\ref{Fig:5} in the rotating frame. The
hyperfine-induced flip-flop transitions of the electron and any
isotope group $\alpha$ of the nuclear spins create non-equilibrium
nuclear spin polarization parallel to the component of the mean spin
of the electrons,  which is static in the rotating frame, with the
rate $S_0/(2 T^{(e)}_{1,N})$ (see Fig.~\ref{Fig:5}). Because of the
nuclear spin precession in constant ($B_e$ and $\tilde
B_{\pm,\alpha}$) and random ($\sqrt{ \xi}B_L$) magnetic fields, only
the nuclear spin projection on the total field $B^*$ (in the
rotating frame) survives. The time-averaged spin flow $q_+$ (or
$q_-$) in the laboratory frame is then obtained by projecting the
spin flow in the rotating frame on the direction of $\tilde
B_{\pm,\alpha}$, which coincides with  $B$. This double projecting
brings about the fraction containing various magnetic fields in
Eq.~\eqref{nuclear_gen1}, which is just a product of the two cosines
arising in the projection procedure. The value of $\xi$ cannot be
obtained from this simple geometric model and requires a microscopic
calculation within the spin-temperature formalism~\cite{OptOr-2,
OptOr-5,Dyakonov75}.

The general rate equation for the nuclear spin projection on the
external field (see Eq.~(11.13) in Ref.~\cite{Kalevich_Ch11}) reads:
\begin{equation}
\langle {\dot I_{\alpha}} \rangle = q_{\alpha} -\frac{1}{T^{(e)}_{1,N}} \left(\frac{\langle I_{\alpha}
\rangle}{Q_{\alpha}} + \langle S_T \rangle \right) -\frac{\langle I_{\alpha} \rangle}{T_N}
\end{equation}
where $Q_{\alpha}=\frac{I_{\alpha}(I_{\alpha}+1)}{s(s+1)}$, $s$ and
$I_{\alpha}$ are the  electron spin and nuclear spin for isotope
$\alpha$. $\langle {S_T}\rangle$   is the  equilibrium value of the
electron spin, corresponding to the thermal population of its Zeeman
sublevels, and $T_N$ is the relaxation time of nuclear spin by all
mechanisms other than interaction with electrons. Assuming that
$\langle {S_T}\rangle \ll S_0 $ and  $T_N \gg T^{(e)}_{1,N}$, we
obtain for the steady state regime:
\begin{eqnarray}
\label{nuclear_eq}
\langle I_{\alpha} \rangle &=& Q_{\alpha}T^{(e)}_{1,N}q_{\alpha}= \nonumber \\
&=&\frac{Q_{\alpha}S_0}{2}\left[\frac{B_{e,\alpha} \tilde B_{+,\alpha}}{(\tilde B_{+,\alpha})^2 + (B_{e,\alpha})^2 + \xi
B_L^2} \right. \nonumber \\
&+& \left. \frac{B_{e,\alpha} \tilde B_{-,\alpha}}{(\tilde B_{-,\alpha})^2 + (B_{e,\alpha})^2 + \xi
B_L^2} \right].
\end{eqnarray}

The hyperfine field created by the mean nuclear spin, acting on the
electron spins is~\cite{OptOr}:
\begin{equation}
\label{bn} B_N = \sum_{\alpha} b_{N, \alpha} \frac{\langle
I_{\alpha} \rangle}{I_{\alpha}} = \sum_{\alpha}
\frac{A_{\alpha}}{\mu_B g_e} \frac{\langle I_{\alpha}
\rangle}{I_{\alpha}}
\end{equation}
where  $ b_{N, \alpha} = A_{\alpha}/(\mu_B g_e)$. The field
$\mathbf{B}_N$ adds to the external field $\mathbf{B}$
($\mathbf{B}_N$ is almost collinear with $\mathbf{B}$) and modifies
the precession frequency of the electron spin $\omega_L=g_e \mu_B
(B+B_N)/\hbar$ in the RSA regime. One can see, that the dependence
of the mean nuclear spin on the external magnetic field (Eq.~(4))
and, therefore, the dependence of the hyperfine field created by
the $\alpha$ isotopes $B_{N,\alpha}$ (Eq.~(5)), has two
derivative-like features (resonances) around $B_{\pm,\alpha}$, i.e.
$B=\pm \omega_\mathrm{m}/\gamma_{\alpha}$. The widths of these
resonances are equal to $\sqrt{(B_{e,\alpha})^2 + \xi B_L^2}$ and
the phases are determined by $b_{N,\alpha}$. Indeed, $A_{\alpha} $
is proportional to the nuclear magnetic moment
$\mu_{\alpha}=\hbar\gamma_{\alpha}$), and, therefore, the Knight
field
$B_{e,\alpha}=b_{e,\alpha}\frac{S_0}{2}=-\frac{A_{\alpha}}{\mu_{\alpha}}\frac{S_0}{2}$
does not depend on the value and sign of $\mu_{\alpha}$. It follows
then from Eq.~(\ref{nuclear_eq}) that the mean nuclear spin gained
by resonant cooling is not sensitive to the sign of  $\mu_{\alpha}$
either. Therefore, the sign of the nuclear field at certain
detunings of the external field from the resonance, according to
Eq.~(\ref{bn}), is determined by the signs of the electron
$g$-factor and the magnetic moment $\mu_{\alpha}$ of the resonant
isotope. For instance, the resonances  in GaAs and CdTe, both having
negative electron $g$-factors, will look inverted  with respect to
each other, because the magnetic isotopes of Cd and Te have negative
hyperfine constants, while for all the nuclear species in GaAs the
hyperfine constants are positive (see Table~\ref{Table 2}). In CdTe
and ZnSe the resonances have the same shape as opposite signs of
their electronic $g$ factors are compensated by the opposite signs
of the hyperfine constants.

\begin{table} [h]
\caption{Parameters of nuclear isotopes with nonzero spin in the studied structures. All data for nuclear spins, $I_{\alpha}$, and magnetic momenta,  $\mu_{\alpha}$, are taken from Ref.~[\onlinecite{Handbook}]. $A_{\alpha}$ is the hyperfine constant, which is taken for a unit cell with two nuclei and $\eta_{\alpha}=| u_c({\bm R}_{\alpha})|^2v_0$, where $u_c({\bm R}_{\alpha})$ is the electron Bloch function at the $\alpha$-th nucleus and $v_0$ is the unit cell volume \cite{syperek11}.}  
\label{Table 2}
\begin{tabular}{p{1.2cm} p{0.8cm}p{1.3cm}p{2cm}p{1.5cm}p{1.5cm}}
\hline
\hline
Species & $I_{\alpha}$ & $\mu_{\alpha}$ & Abundance $\varkappa_{\alpha}$ & $\eta_{\alpha}$ ($\times 10^3$) & $A_{\alpha}$  ($\mu$eV) \\
\hline
$^{111}$Cd & 1/2 & -0.5943 & 0.128 & 3.6 \footnote[1]{From Ref.~[\onlinecite{Nakamura79}]} & -37.4 \\
$^{113}$Cd & 1/2 & -0.6217 & 0.123 & 3.6 \footnotemark[1] & -39.1 \\
\\
$^{125}$Te & 1/2 & -0.8871 & 0.079 &  & -45 \footnote[2]{From Ref.~[\onlinecite{Testelin09}]}   \\
\\
$^{67}$Zn & 5/2 & +0.8754 & 0.041 &  & 3.7 \footnote[3]{From Ref.~[\onlinecite{Greilich12}]}  \\
\\
$^{77}$Se & 1/2 & +0.534 & 0.0758 & 3.6 & 33.6\footnote[4]{From Ref.~[\onlinecite{syperek11}]}\\
\\
$^{69}$Ga & 3/2 & +2.016 & 0.604 & 2.61 \footnote[5]{From Ref.~[\onlinecite{Paget77}]} & 38.2 \\
$^{71}$Ga & 3/2 & +2.562 & 0.396 & 2.61 \footnotemark[5] & 48.5 \\
\\
$^{75}$As & 3/2 & +1.439 & 1 & 4.42 \footnotemark[5] & 46\\
\hline
\end{tabular}
\end{table}

\subsection{NMR in RSA spectra}
\label{sec:6}

In the RSA spectra the electron spin polarization is measured via
the Kerr rotation. Information on the nuclear spin polarization and
NMR can be also obtained from the RSA spectra when the nuclei modify
the electron spin polarization. To model nuclear effects in RSA
spectra we use a two-step procedure.  In the first step, we neglect
$B_N$, because this field is weak  in the resonant-cooling regime
and calculate the electron spin polarization (and therefore the
Knight field) under light excitation with alternated helicity in the
external magnetic field. For these calculations we need to account
for: (i) the periodic change of the pump polarization (caused by the
PEM); (ii) the  generation of  electron spin polarization by the
short pump pulse; (iii) the electron spin dynamics in the magnetic
field; and (iv) the accumulation of the electron spin polarization
after a train of pump pulses. This would allow us to calculate the
hyperfine field, which develops as a result of resonant cooling in
the sum of the external magnetic field and the Knight field. In  the
second step, we calculate the electron spin polarization, taking
into account the nuclear field, and model the resulting RSA spectra.

\subsubsection{Polarization of the pump pulses}

One of the main features of the RSA technique is the periodic generation
of the electron spin coherence by a train of laser pulses. In our experiments the
repetition  period of the laser pulses is $T_R=13.2$~ns (the pump pulse repetition
frequency is 75.7~MHz). The polarization of the  pump pulses is changed
periodically by the PEM with a frequency $f_\mathrm{m} = 42 \div 168$~kHz.

We model the excitation with light of alternating  helicity in the
following way. Let the pump beam before the PEM be polarized
linearly along the  $x$ axis: $\bm E(\bm r, t) = E_{x}(\bm r,t) \bm
o_x + c.c.$, here $E_x \sim \mathrm e^{-\mathrm i \omega t}$,
$\omega$ is the optical frequency,  and $\bm o_x$ is the unit vector
along $\mathbf{x}$. The electric vector of the light after the PEM
can be written as:
\begin{eqnarray}
\label{eq:pem_all} \bm E(\bm r, t) = \frac{E_{x}(\bm r,t)}{\sqrt{2}}
\left( \cos \left[\frac{\phi}{2} \sin(\omega_\mathrm{m} t)- \frac{\pi}{4}
\right ] \bm o_+ \right.
\\ \nonumber
 \left.+ \sin \left[ \frac{\phi}{2} \sin(\omega_\mathrm{m} t)- \frac{\pi}{4}  \right ] \bm o_- \right)+ {\rm c.c.}\:,
\end{eqnarray}
where $\bm o_\pm$ are the circularly polarized unit vectors that are
related to the unit vectors ${\bm o}_x \parallel \mathbf{x}$ and ${\bm o}_y
\parallel \mathbf{y}$ through $\bm o_\pm = (\bm o_x \pm \mathrm i \bm
o_y)/\sqrt{2}$. $\phi$ is the maximal phase delay created by PEM. In
our Kerr rotation experiments $\phi=\pi/2$, when we use the first
harmonic. One can expand  Eq.~\eqref{eq:pem_all} into a harmonic
series:
\begin{widetext}
\begin{eqnarray}
\label{eq:pem_series}
&&\bm E(\bm r, t) = \frac{E_{x}(\bm r,t)}{2}
\left[\left[ -J_0 \left(\frac{\pi}{4}\right)+
2\sum_{k=0}^\infty \left(J_{2k} \left(\frac{\pi}{4}\right) \cos(2k\omega_\mathrm{m} t) + J_{2k+1} \left(\frac{\pi}{4}\right)
\sin((2k+1)\omega_\mathrm{m} t) \right) \right] \right. \bm o_+  \\ \nonumber +
&&\left. \left[ J_0 \left(\frac{\pi}{4}\right) -
2\sum_{k=0}^\infty \left( J_{2k} \left(\frac{\pi}{4}\right)
\cos(2k\omega_\mathrm{m} t) - J_{2k+1} \left(\frac{\pi}{4}\right)
\sin((2k+1)\omega_\mathrm{m} t) \right) \right] \bm o_-\right]  +  {\rm c.c.}\:,
\end{eqnarray}
\end{widetext}
where $J_k$ are Bessel functions.

Choosing $\phi = \pi/2$ and neglecting all terms that oscillate at
higher harmonic frequencies ($2\omega_\mathrm{m}$,
$3\omega_\mathrm{m}$, \ldots) we can approximately rewrite
Eq.~\eqref{eq:pem_series} as:
\begin{eqnarray}
\label{eq:pem} \bm E(\bm r, t) = \frac{E_{x}(\bm r,t)}{2} \left(
\left[ 2J_{1} \left(\frac{\pi}{4}\right) \sin(\omega_\mathrm{m} t)-J_0
\left(\frac{\pi}{4}\right) \right] \bm o_+\right. \\ \nonumber
 + \left. \left[ 2J_{1} \left(\frac{\pi}{4}\right) \sin(\omega_\mathrm{m} t)+J_0 \left(\frac{\pi}{4}\right) \right] \bm o_- \right) + {\rm c.c.} \\ \nonumber
 \equiv E_{\sigma^{+}}(\bm r, t)\bm o_++E_{\sigma^{-}}(\bm r, t)\bm o_- + {\rm c.c.}\:.
\end{eqnarray}
The modulation of the light by the PEM generates elliptically
polarized light that  changes from $\sigma^+$ to $\sigma^-$ going
through linear}.

\subsubsection{Generation of electron spin polarization}

Our consideration of the generation of the spin polarization for the
resident electrons in QWs and the electrons bound to donors in bulk
semiconductors is based on the approach developed in
Refs.~\onlinecite{zhu10,rsa_paper,Yugova12}. We assume that the
long-lived spin polarization of the electrons  is generated by
resonant excitation of the negatively charged excitons (trions).
Below, we consider the case of trions in QWs, while the approach is
equally valid for the resonant excitation of the donor-bound
excitons in bulk semiconductors. The resonant excitation of the
electron spin system by a short elliptically polarized pulse  can be
described as:
\begin{eqnarray}
\label{eq:exc}
S_z^a=\frac{Q_+^2-Q_-^2}{4}+\frac{Q_+^2+Q_-^2}{2}S_z^b \\ \nonumber
S_y^a=Q_+Q_-S_y^b,\\ \nonumber
S_x^a=S_x^b.
\end{eqnarray}
Here $S_z$, $S_y$, $S_x$ are electron spin components, the subscript
$a$ and $b$ denote the spin components at a time just after or
shortly before the pump pulse arrival. In the magnetic field
$\textbf{B}||\textbf{x}$ the $S_x$ component does not change and  we
disregard it in the following. $Q_+$ and $Q_-$ are associated with
the powers of the $\sigma^+$  and $\sigma^-$ components of the
incident light. They are defined by the corresponding pulse areas
$\Theta_{\pm}$: $Q_{\pm}=\cos{(|\Theta_{\pm}|/2)}$, $\Theta_{\pm} =
\int 2\langle d \rangle E_{\sigma^{\pm}}(t)dt / \hbar$. Here
$\langle d \rangle$ is the dipole transition matrix element and
$E_{\sigma^{\pm}}(t)$ are smooth envelopes of the circular
components of the electric field of the laser pulse.

It follows from Eq.~\eqref{eq:pem} that
$E_{\sigma^{\pm}}(t)=E_{x}(\bm r,t) \left[ J_{1}
\left(\frac{\pi}{4}\right) \sin(\omega_\mathrm{m} t) \mp J_0
\left(\frac{\pi}{4}\right)/2 \right]$ and $\Theta_{\pm} \approx
\Theta_0\left[ J_{1}
\left(\frac{\pi}{4}\right)\sin(\omega_\mathrm{m} t) \mp J_0
\left(\frac{\pi}{4}\right)/2 \right]$, where $\Theta_0=\int \langle
d \rangle E_x(t)dt / \hbar$. In the low pump power regime
($|\Theta_0|^2 \ll 1$), which is valid for QWs \cite{rsa_paper},
Eq.~\eqref{eq:exc} gives us:
\begin{eqnarray}
\label{eq:exc_low}
S_z^a \approx \frac{|\Theta_0|^2}{16}J_{0}\left(\frac{\pi}{4}\right) J_{1}\left(\frac{\pi}{4}\right) \sin(\omega_\mathrm{m} t)+S_z^b \\
\nonumber S_y^a \approx S_y^b.
\end{eqnarray}

\subsubsection{Spin dynamics in a magnetic field}

Since the time scales of the electron and nuclear spin dynamics
differ by several orders of magnitude, the fast dynamics of the
single electron spin in the interval between pump pulses occurs in
the sum of the external magnetic field and the frozen effective
field of the nuclear spins, $\mathbf{B} + \mathbf{B}_N$.
\begin{multline}
\label{eq:prec} \mathbf{S}(t) = [ \mathbf{n} (\mathbf{n} \cdot \mathbf{S}^{a} ) + (\mathbf{S}^{a} - \mathbf{n} (\mathbf{n}\cdot \mathbf{S}^{a} )) \cos{(\omega_L t)} \\ + [\mathbf{S}^{a} - \mathbf{n} (\mathbf{n} \cdot \mathbf{S}^{a} )]\times \mathbf{n} \sin{(\omega_L t)} ] \exp{(-t/T_2)},
\end{multline}
where $\mathbf{n} =  (\mathbf{B}+ \mathbf{B}_N)/|\mathbf{B}+
\mathbf{B}_N|$ is the unit vector along the total magnetic field,
$\omega_L = |g_e \mu_B (\mathbf{B}+ \mathbf{B}_N)|/\hbar$ is the
electron Larmor precession frequency, and $T_2$ is the transverse
spin relaxation time of electrons (i.e. electron spin coherence
time). It  is worth reminding that at the first step we neglect
$\mathbf{B}_N$ for the calculation of $\mathbf{S}(t)$.

The modulation period of the pump polarization is three orders of
magnitude longer than the laser repetition period $T_R$.  This leads
to accumulation of electron spin polarization after a train of pump
pulses with a polarization determined by the PEM.

\subsubsection{Accumulation of electron spin polarization and Knight field}

The accumulated electron spin polarization after each repetition
period, $\mathbf{S}(T_R)$ (given by Eq.~\eqref{eq:prec}) should be
equal to the spin right before the pump pulse arrival,
$\mathbf{S}^b$ (given by Eqs.~\eqref{eq:exc}). Without an effective
nuclear field, the resulting electron spin polarization after each
pump pulse is
\begin{eqnarray}
\label{eq:rsa_ell}
S_z^a &=& -\frac{Q_+^2-Q_-^2}{4\Delta}[K_1\cos(\omega_L T_R)-1], \\
\nonumber S_y^a &=& -\frac{Q_+^2-Q_-^2}{4\Delta}K_1\sin(\omega_L T_R),\\
\nonumber \Delta &=& 1-(K_1+K_2)\cos(\omega_L T_R)+K_1K_2, \\
\nonumber K_1 &=& Q_+Q_- \exp(-T_R/T_2), \\
\nonumber K_2 &=&  \frac{Q_+^2+Q_-^2}{2}\exp(-T_R/T_2).
\end{eqnarray}

\begin{figure}[hbt]
\includegraphics*[width=8cm]{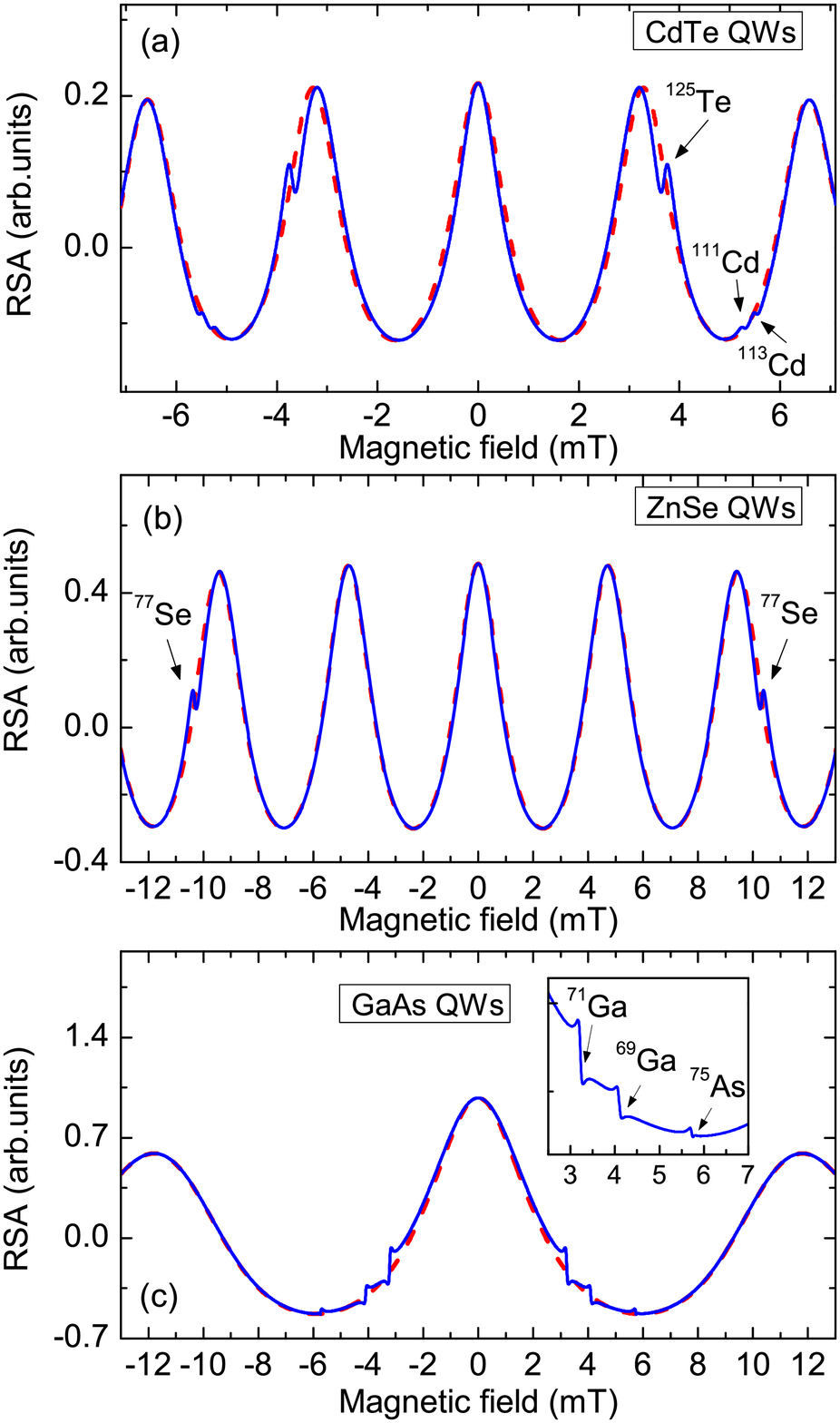}
\caption[] {(Color online) Calculated RSA spectra for CdTe-, ZnSe- and
GaAs-based QWs. Dashed (red) lines show RSA spectra without accounting for
the nuclear effects. Solid (blue) lines show RSA spectra with
the nuclear spin effects taken into account.}
\label{Fig:6}
\end{figure}

The experimental RSA signal is proportional to the $S_z$ component
of the  accumulated polarization taken in one of the circular
polarizations $\sigma^+$ or $\sigma^-$ (e.g., for $\sigma^+$
polarization $\Theta_-=0$ and, therefore, $Q_-=1$). Modeled  RSA
spectra without taking into account nuclear spin effects are shown
in Fig.~\ref{Fig:6} (red curves) for three material systems based on
CdTe, ZnSe and GaAs. For the electron spin ensemble the calculated
polarization is averaged over the $g$-factor spread with a Gaussian
distribution function (see Sec.III.D in Ref.~\onlinecite{Yugova12}).
The parameters used for these calculations are given in
Table~\ref{Table 3}. The electron spin coherence time $T_2$ and the
$g$-factor spread $\Delta g_e$ were taken close to the experimental
data.

We now turn to the calculation of the Knight field. Here we
consider again a single electron spin. Note, that as the Bohr
magneton, $\mu_\mathrm{B}$, is approximately 2000 times larger than
the nuclear magneton, the Zeeman splitting of the electron and
nuclear spin levels and their Larmor precession frequencies differ
by three orders of magnitude. Therefore, the slow dynamics of the
NSS is determined only by the averaged electron spin polarization.
For the average of the polarization over the period from the
$(m-1)$th to the $m$th pulse, we obtain
\begin{multline}
\label{s0}
\mathbf{S}_0 = \frac{1}{T_R} \int_{(m-1)T_R}^{mT_R}\mathbf{S}(t)  \mathrm dt~ =\\
\mathbf{n} (\mathbf{n} \cdot \mathbf{S}^a) + \frac{\mathbf{S}^a - \mathbf{n} (\mathbf{n}\cdot \mathbf{S}^a )}{\omega_L T_R} \sin{(\omega_L T_R)} \\ +
\frac{[\mathbf{S}^a - \mathbf{n} (\mathbf{n} \cdot \mathbf{S}^a )]\times \mathbf{n}}{\omega_L T_R} [1-\cos{(\omega_L T_R)}].
\end{multline}
Taking into account Eq.~\eqref{eq:rsa_ell} we can write:
\begin{eqnarray}
\mathbf{S}_0&=&\sqrt{\frac{[(S_z^a)^2+(S_y^a)^2]K_3}{(T_R/T_2)^2+(\omega_L
T_R)^2}}, \\ \nonumber K_3&=&\sqrt{(1-2\exp(-T_R/T_2)\cos(\omega_L
T_R)+\exp(-2T_R/T_2))}.
\end{eqnarray}
For QWs all these equations are valid in the low-power regime
\cite{rsa_paper}. This gives us:
\begin{equation}
\mathbf{S}_0 \approx -\frac{|\Theta_0|^2}{16}J_{0}\left(\frac{\pi}{4}\right) J_{1}\left(\frac{\pi}{4}\right) \frac{\sin (\omega_\mathrm{m} t)}{\sqrt{(T_R/T_2)^2+(\omega_L T_R)^2}}.
\end{equation}
This is the specific expression for the oscillating electron spin
polarization of Eq.~\eqref{nuclear_gen}, which results in NMR lines
in the RSA spectra.

\subsubsection{RSA spectra with nuclear spin effects}

We use equation (15) for the electron field $B_e=b_e S_0/2$ to
calculate from Eqs.~\eqref{nuclear_eq} and \eqref{bn} the hyperfine
field $B_N$. Then, at the second step of our calculations, we take
again Eqs.~\eqref{eq:exc} and \eqref{eq:prec} and calculate the
accumulated electron spin polarization in the sum of external and
nuclear magnetic fields, similar to Eq.~\eqref{eq:rsa_ell}. The
calculated polarization is again averaged over the $g$-factor spread
with a Gaussian distribution function.

Modeled RSA spectra including the effect of the nuclear field are
shown in Fig.~\ref{Fig:6} by the blue curves. The parameters for the
calculations are given in Table~\ref{Table 3}. $b_e$ and
$b_{N,\alpha}$ were evaluated using the $A_{\alpha}$ and
$\mu_{\alpha}$ values from Table~\ref{Table 2}. The parameter value
$\sqrt{\xi} B_L=0.05$~mT controlling the width of the NMR features was
taken identical for all materials.

\begin{table} [hbt]
\caption{Parameters used for calculation of RSA spectra in
Fig.~\ref{Fig:6}.} \label{Table 3}
\begin{tabular}{p{1.2cm} p{2.3cm}p{2.7cm}p{2.3cm}}
\hline \hline
 QWs & CdTe & ZnSe & GaAs \\
\hline
$g_e$ &   $-$1.64    &   1.13           & $-$0.44   \\
$\Delta g_e$ & 0.04 &0.02  &  0.055  \\ \\
$b_e$& $-$43 ($^{111}$Cd) &$-$13 ($^{67}$Zn) &$-$2.6 ($^{69}$Ga) \\
(mT)  & $-$43 ($^{113}$Cd) &$-$19.4 ($^{77}$Se) &$-$2.6 ($^{71}$Ga) \\
 & $-$54 ($^{125}$Te)& &$-$4.3 ($^{75}$As) \\ \\
$b_{N,\alpha}$ & 52 ($^{111}$Cd)& 2.4 ($^{67}$Zn) &$-$900 ($^{69}$Ga)\\
(mT)  & 52 ($^{113}$Cd)&20 ($^{77}$Se) &$-$760 ($^{71}$Ga) \\
    & 61 ($^{125}$Te)& & $-$1800 ($^{75}$As)\\ \\
    $f_m$ & 50 & 84 &42  \\
(kHz)& & & \\ \\
$T_2$(ns) & 10.6 & 10.6& 13.2  \\ \\
\hline
\end{tabular}
\end{table}

Comparing the blue and red graphs in Fig.~\ref{Fig:6} one can
conclude about three effects related to the electron-nuclear
interaction, which become evident in the RSA spectra: (i) Appearance
of NMR features for different isotopes; (ii) Narrowing (in CdTe and
ZnSe) or broadening (in GaAs) of the zero-field RSA peak; and (iii)
Shift of the nonzero-field RSA peaks, which is pronounced for CdTe.
One can see, that the phase of the derivative shape of the NMR
features in CdTe and ZnSe coincide with each other, while it is
inverted in GaAs. The phase is controlled by the sign of the
$b_{N,\alpha}$ field (see Table~\ref{Table 3}), which either
increases or decreases the equilibrium polarization of the nuclei
induced by the external field.  The same reason explains the changes
in the linewidth of the zero-field RSA peak in the effect (ii). The
shift of the non-zero RSA peaks (effect (iii)) is related to the
developing of the $B_{N,x}$ component of the DNP. Its experimental
appearance and theoretical description will be published elsewhere.

\section{Conclusions}\label{sec:conclusions}

Our experiments and their comparison with the developed theoretical
model have demonstrated the possibility to realize resonant cooling
of the nuclear spin system in semiconductor nanostructures and bulk
layers in the regime of resonant spin amplification (RSA). This was
done by thorough adjustment of the magnetic field to the exact Voigt
geometry and using a photoelastic modulator to alternate the
helicity of the excitation laser at high frequency, keeping the
time-averaged spin of optically oriented electrons zero within the
experimental precision. Under these conditions, nuclear spin pumping
is forbidden outside the resonance of the nuclear Larmor precession
with the modulation frequency. All-optical NMR signals with
dispersion-like shape, typical for resonant cooling, were observed
when the NMR resonance field was close to the position of the RSA
peaks. The NMR resonances were detected on up to the fifth RSA peak.
Thus, resonant cooling signals were observed at magnetic fields well
outside the typical width of Hanle curves for the studied
structures, which are nearly identical to the shape of the
zero-field RSA peak~\cite{comment_hanle}.

As follows from the developed theory, these signals are due to
cooling of the nuclei by both the $z$ and $y$-components of the
electron mean spin. Their vector sum decays with growing transverse
field $B$ as $1/B$ rather than $1/B^2$ as does the $z$-component
usually observed in the Hanle effect. Sharp RSA peaks that rise at
the magnetic fields satisfying electron spin resonance conditions at
the laser repetition frequency and its overtones, provide the probe
for the nuclear spin polarization via its effect upon the electron
spin resonance frequency and help detecting resonant cooling signals
at magnetic fields well outside the width of the Hanle curve. This
way, we detected all-optical NMR of all the magnetic isotopes
present in the studied samples. Notably, the shape of the
experimentally observed resonances are sensitive to the sign of the
isotope magnetic moments, in full agreement with the theory. The
difference of our findings with previously reported  experiments on
all-optical NMR in the RSA regime (for example, \cite{Sal01a}) is
that in our case there was no background nuclear polarization, and
all observed nuclear-spin effects developed exclusively at resonance
fields due to resonant cooling of the nuclear spin system.

\acknowledgments

The authors are thankful to I. A. Merkulov for valuable discussions.
This work was supported by the Deutsche Forschungsgemeinschaft, BMBF
(project 05K12PE1), the Russian Ministry of Education and Science (contract No.11.G34.31.0067 with SPbSU and leading scientist A. V. Kavokin), and EU FET project SPANGL4Q.


\begin{thebibliography}{34}
\bibitem{comment_hanle}
The zero-field RSA peak [Eq.~\eqref{eq:rsa_ell}] has the same form as
the standard expression for the Hanle effect \cite{OptOr}: the
electron spin depolarization in a transversal magnetic field under
continuous wave pumping. The influence of the inhomogeneous
distribution of $g$ factors on the Hanle effect is typically quite
weak and, as a rule~\cite{ast08,Yugova12}, the extracted spin
dephasing time is controlled by the nuclear spin fluctuations, i.e.,
by $T_2^*$. As compared with the Hanle effect, the studies of the
resonant spin amplification allow one to directly extract the
magnetic field dependence of the spin dephasing time, and
consequently, evaluate the spread of $g$ factors, $\Delta g$.
\end{thebibliography}
\end{document}